\documentclass[aps,pre,superscriptaddress,twocolumn,showpacs]{revtex4}
\usepackage{epsfig}
\usepackage{graphicx}
\usepackage{dcolumn}
\usepackage{bm}
\usepackage{amsmath,amssymb,times}

\begin{document}

\title{Kuramoto dilemma alleviated by optimizing connectivity and rationality}

\author{Han-Xin Yang}\email{hxyang01@gmail.com}
\affiliation{Department of Physics, Fuzhou University, Fuzhou
350116, People's Republic of China}

\author{Tao Zhou}\email{zhutou@ustc.edu}
\affiliation{Institute of Fundamental and Frontier Sciences,
University of Electronic Science and Technology of China, Chengdu
611731, People¡¯s Republic of China} \affiliation{CompleX Lab, Web
Sciences Center, University of Electronic Science and Technology of
China, Chengdu 611731, People's Republic of China} \affiliation{Big
Data Research Center, University of Electronic Science and
Technology of China, Chengdu 611731, People's Republic of China}

\author{Zhi-Xi Wu}\email{eric0724@gmail.com}
\affiliation{Institute of Computational Physics and Complex Systems,
Lanzhou University, Lanzhou, Gansu 730000, People's Republic of
China}

\begin{abstract}

Recently, Antonioni and Cardillo proposed a coevolutionary model
based on the intertwining of oscillator synchronization and
evolutionary game theory [Phys. Rev. Lett. \textbf{118}, 238301
(2017)], in which each Kuramoto oscillator can decide whether to
interact-or not-with its neighbors, and all oscillators can receive
some benefits from the local synchronization but those who choose to
interact must pay a cost. Oscillators are allowed to update their
strategies according to payoff difference, wherein the strategy of
an oscillator who has obtained higher payoff is more likely to be
followed. Utilizing this coevolutionary model, we find that the
global synchronization level reaches the highest level when the
average degree of the underlying interaction network is moderate. We
also study how synchronization is affected by the individual
rationality in choosing strategy.

\end{abstract}

\date{\today}

\pacs{89.75.Hc, 05.45.Xt, 02.50.Le}

\maketitle

\section{Introduction}

Synchronization is the coordination of events to operate a system in
unison, which is ubiquitous in natural world and human society. For
example, two pendulum clocks suspended side by side swing with the
same frequency, a school of fish move in the same
direction~\cite{vicsek}, opinions of different individuals reach a
consensus~\cite{opinion}, all generators in a power-grid system
rotate at the same frequency~\cite{power}, and so on. Due to the
rapid development of network
science~\cite{barabasi,newman,Boccaletti}, the study of
synchronization in complex networks has attracted increasing
attention~\cite{review1,review2}. It has been found that both the
network structure~\cite{network1,network2,network3} and the coupling
scheme~\cite{coupling1,coupling2,coupling3} play crucial roles in
the synchronization processes.

Among many models for synchronization phenomena, the Kuramoto model
is the most popular nowadays~\cite{kuramoto1,kuramoto2,kuramoto3},
where many oscillators with heterogeneous natural frequencies couple
through the sine of their phase differences. When this coupling is
strong enough, all oscillators eventually rotate at the same
frequency. In its original model, all agents interact with their
neighbors all the time. However, the interaction is usually costly
in reality. Due to the existence of costly interactions, some agents
may choose not to interact with neighbors at some time, which
impedes synchronization of the whole system.

How to understand the emergence of synchronization with costly
interactions is a challenging issue. Recently, Antonioni and
Cardillo incorporated the evolutionary game in the Kuramoto
model~\cite{prl}. They assumed that agents can decide whether to
interact-or not-with their neighbors. All agents receive the benefit
from the local synchronization but those who choose to interact with
their neighbors must pay some cost. Each agent's payoff is thus
given by the net gain of benefit and cost. After each time step,
agents are allowed to update their strategies by comparing payoffs
with their neighbors. The strategies of agents with higher payoffs
are more likely to be imitated. In this sense, this toy model can be
named as the evolutionary Kuramoto game (EKG).

The EKG sheds some light on the study of self-organized
interactions. Agents in EKG face a dilemma, that is, the interaction
with neighbors helps to enhance synchronization of the whole system
but the interaction cost weakens the willingness of agents'
participation. So far, how to alleviate this dilemma and enhance the
global level of synchronization in EKG are still unclear. In this
paper, we find that both too few and too many neighbors would
suppress synchronization in EKG. Besides, we have shown that the
rationality of agents plays a non-trivial role in EKG.

\section{Model}

Let us consider a network of $N$ nodes and average degree $\langle k
\rangle$. Each node $i$ represents a Kuramoto oscillator and it can
decide whether to interact-or not-with its neighbors. We call an
agent as a cooperator if it interacts with its neighbors. Those who
do not participate in such interaction are called as defectors.

Following the work of Antonioni and Cardillo~\cite{prl}, agent $l$
changes its phase $\theta_{l}$  according to the following equation:
\begin{equation}
\dot{\theta_{l}}=\omega_{l}+s_l\lambda\sum_{m=1}^{N}a_{lm}\sin(\theta_{m}-\theta_{l}),\label{1}
\end{equation}
where $\omega_{l}$ is the natural frequency of $l$, $s_l$ is the
strategy of $l$ ($s_l=1$ if $l$ is a cooperator and $s_l=0$ if it is
a defector), $\lambda$ is the coupling strength, and $a_{lm}$ is the
element of the adjacency matrix ($a_{lm}=1$ if $l$ and $m$ is
connected, and $a_{lm}=0$ otherwise). In this paper, both the
initial phases $\theta$ and the natural frequencies $\omega$ of all
agents are uniformly distributed over the interval [-$\pi$,$\pi$].

The local order parameter for a pair of nodes $l$ and $m$, $r_{lm}$,
is calculated by
\begin{equation}
r_{lm}e^{i(\theta_{l}+\theta_{m})/2}=\frac{e^{i\theta_{l}}+e^{i\theta_{m}}}{2}.\label{2}
\end{equation}
The benefit of agent $l$ is defined as its local order parameter,
which is determined by
\begin{equation}
r_{l}=\frac{\sum_{m=1}^{N}a_{lm}r_{lm}}{\sum_{m=1}^{N}a_{lm}}.
\label{3}
\end{equation}
The cost of an agent $l$ is defined as the absolute value of the
angular acceleration, i.e.,
\begin{equation}
c_{l}=|\dot{\theta_{l}}(t)-\dot{\theta_{l}}(t-\varepsilon)|,
\label{4}
\end{equation}
where $\varepsilon$ is the step length used to compute Eq.~(\ref{1})
with the fourth-order Runge-Kutta method. Note that the cost for a
defector is zero since its frequency is fixed at the natural
frequency. Finally, the payoff of an agent is defined as
\begin{equation}
P_{l}=r_{l}-\alpha\frac{c_{l}}{2\pi}, \label{5}
\end{equation}
where the cost is divided by $2\pi$ to make it commensurable with
the benefit and a scalar $\alpha$ is named as the relative cost.

Initially, cooperators and defectors are randomly distributed in a
network with equal probability. All agents synchronously update
their strategies at discrete time steps $t_{k}$
($t_{k+1}-t_{k}=\varepsilon$) according to the famous Fermi
rule~\cite{fermi}. Specifically, an agent $l$ randomly selects one
of its neighbors, $m$, and adopts its strategy with a probability
given by
\begin{equation}\label{6}
W(s_{l}\leftarrow s_{m})=\frac{1}{1+\exp[-\beta(P_m-P_l)]},
\end{equation}
where the parameter $\beta$ ($>0$) characterizes noise to permit
irrational choices (henceforth we call $\beta$ as the rationality
parameter). Agents become more rational, i.e., have greater
probability to follow the strategies of neighbors who have obtained
higher payoffs, as the rationality parameter $\beta$ increases.

After a sufficiently long transient time, one can calculate the
average level of cooperation $\langle C \rangle$ and the global
synchronization $\langle r_{G} \rangle$. Here $\langle C \rangle$ is
defined as the fraction of cooperators in the steady state. The
global order parameter $r_{G}$ is calculated by
\begin{equation}
r_{G}e^{i\Psi}=\frac{1}{N}\sum_{m=1}^{N}e^{i\theta_{m}},\label{7}
\end{equation}
where $i$ is the imaginary unit and $\Psi$ is the average phase of
the system.

\section{Analyses}

Without loss of generality, we carry out our study of the model in
Erd\H{o}s-R\'{e}nyi (ER) random graphs~\cite{er} with size $N=2000$.
We set the step length $\varepsilon=0.01$ in this paper (we have
checked that qualitative results keep unchanged for a wide range of
$\varepsilon$). In all simulations, we first wait $90 000$ Monte
Carlo time steps to let the system attain steady state, and then run
another $10 000$ Monte Carlo time steps to calculate the average
level of cooperation (synchronization) $\langle C \rangle$ ($\langle
r_{G} \rangle$). Finally, each data presented below results from an
average over 200 independent realizations.

\subsection{Effects of the relative cost $\alpha$ and the coupling strength $\lambda$}

\begin{figure}
\begin{center}
 \scalebox{0.4}[0.4]{\includegraphics{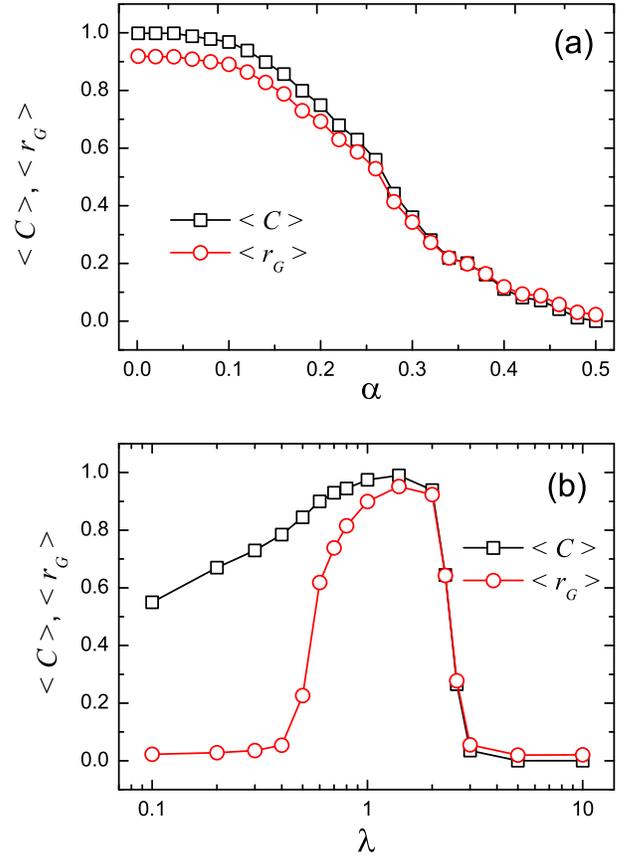}} \caption{(Color online)
The average level of cooperation (synchronization) $\langle C
\rangle$ ($\langle r_{G} \rangle$) as a function of (a) the relative
cost $\alpha$ and (b) the coupling strength $\lambda$, respectively.
The average degree of the network $\langle k \rangle=8$ and the
rationality parameter $\beta=1$. For (a), the coupling strength
$\lambda=1$. For (b), the relative cost $\alpha=0.1$.} \label{fig1}
\end{center}
\end{figure}

\begin{figure*}
\begin{center}
 \scalebox{0.8}[0.8]{\includegraphics{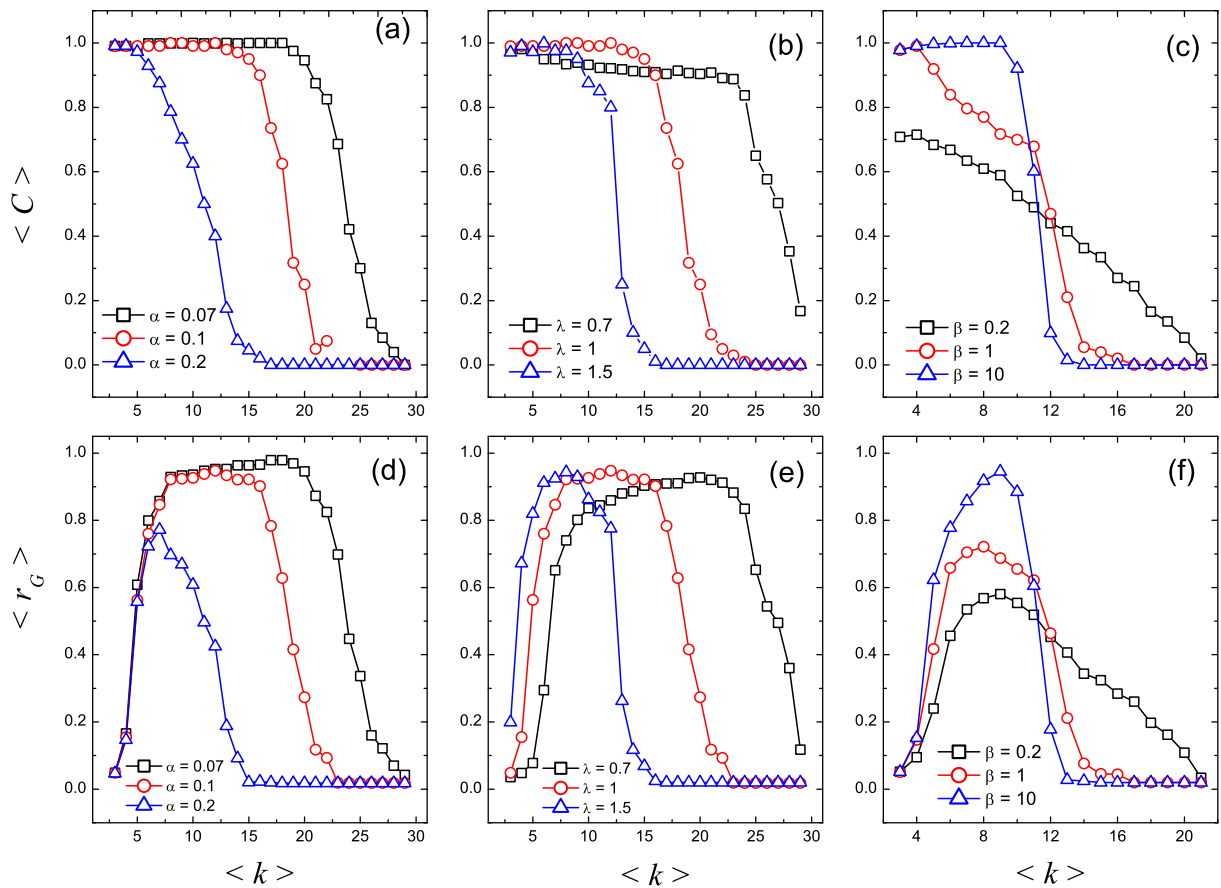}}
 \caption{(Color online) The average level of cooperation $\langle C
\rangle$ as a function of the average degree $\langle k \rangle$ for
different values of (a) the relative cost $\alpha$, (b) the coupling
strength $\lambda$ and (c) the rationality parameter $\beta$,
respectively. The average level of synchronization $\langle r_{G}
\rangle$ as a function of the average degree $\langle k \rangle$ for
different values of (d) the relative cost $\alpha$, (e) the coupling
strength $\lambda$ and (f) the rationality parameter $\beta$,
respectively. For (a) and (d), the coupling strength $\lambda=1$ and
the rationality parameter $\beta=1$. For (b) and (e), the relative
cost $\alpha=0.1$ and the rationality parameter $\beta=1$. For (c)
and (f), the relative cost $\alpha=0.2$ and the coupling strength
$\lambda=1$.} \label{fig2}
\end{center}
\end{figure*}

We first briefly review the effects of the relative cost $\alpha$
and the coupling strength $\lambda$ on the emergence of
synchronization and cooperation, which reproduce the findings
reported in~\cite{prl}. Figure~\ref{fig1} shows the average level of
cooperation (synchronization) $\langle C \rangle$ ($\langle r_{G}
\rangle$) as a function of the relative cost $\alpha$ and the
coupling strength $\lambda$. One can find that $\langle C \rangle$
is positively correlated with $\langle r_{G} \rangle$, that is, the
more fraction of cooperators, the higher level of the
synchronization. From Fig.~\ref{fig1}(a), one can observe that both
$\langle C \rangle$ and $\langle r_{G} \rangle$ decrease as the
relative cost $\alpha$ increases. Obviously, the increase of
$\alpha$ leads to more expensive interaction, which decreases the
payoffs of cooperators. As a result, cooperators are more likely to
be replaced by defectors. The dependence of $\langle C \rangle$
($\langle r_{G} \rangle$) on the coupling strength $\lambda$
exhibits a non-monotonic phenomenon, i.e., $\langle C \rangle$
($\langle r_{G} \rangle$) reaches the maximum when the value of
$\lambda$ is moderate (see Fig.~\ref{fig1} (b)). There exists two
critical values of $\lambda$: $\lambda_{c1}$ and  $\lambda_{c2}$.
For $\lambda<\lambda_{c1}$ or $\lambda>\lambda_{c2}$, $\langle r_{G}
\rangle \simeq0$.

The lower limit $\lambda_{c1}$ is determined by the network
structure as
\begin{equation}
\lambda_{c1}=\lambda_{MF}\frac{\langle k \rangle}{\langle k
^{2}\rangle}, \label{8}
\end{equation}
where $\lambda_{MF}$ ($\lambda_{MF}$=4 in this paper) is the
critical value of the coupling strength in the mean field
case~\cite{mf}. Below $\lambda_{c1}$, oscillators rotate
incoherently even they are all cooperators. From Eqs.~(\ref{1})
and~(\ref{4}), one can note that the interaction cost is positively
related to the coupling strength $\lambda$. Too large value of
$\lambda$ tremendously reduces the ratio of benefit to cost, leading
to the extinction of cooperators and an incoherent state of the
system. According to~\cite{prl}, cooperators can survive when
\begin{equation}
\frac{\sqrt{2+2sin(\varepsilon \lambda) }-\sqrt{2}
}{\varepsilon\lambda \langle k \rangle}\pi>\alpha. \label{9}
\end{equation}
To satisfy the above equation, the coupling strength must be below a
certain value $\lambda_{c2}$.

We want to emphasize that the critical values of phase transitions
are accurate for large system size. But in this paper, we do not
focus on the phase transitions and hence use a small system.

\subsection{Effects of the average degree $\langle k \rangle$}

Figure~\ref{fig2} shows the average level of cooperation $\langle C
\rangle$ (synchronization $\langle r_{G} \rangle$) as a function of
the average degree $\langle k \rangle$ for different values of the
relative cost $\alpha$, the coupling strength $\lambda$ and the
rationality parameter $\beta$. From Fig.~\ref{fig2}
(a)-\ref{fig2}(c), we observe that the cooperation level $\langle C
\rangle$ keep almost unchanged for small values of $\langle k
\rangle$ and then decreases to zero as $\langle k \rangle$
continually increases.

In the original Kuramoto model, the increase of the average degree
facilitates the attainment of synchronization~\cite{degree}.
However, the dependence of $\langle r_{G} \rangle$ on $\langle k
\rangle$ in the EKG displays a nonmonotonic behavior. As shown in
Fig.~\ref{fig2} (d)-\ref{fig2}(f), there exists an optimal value of
$\langle k \rangle$, at which $\langle r_{G} \rangle$ is maximized.
From Fig.~\ref{fig2}(d) and \ref{fig2}(e), we observe that the
optimal value of $\langle k \rangle$ tends to decrease as the
relative cost $\alpha$ or the coupling strength $\lambda$ increases.
For $\alpha$ = 0.07, 0.1 and 0.2, the optimal value of $\langle k
\rangle$ is 18, 12 and 7 respectively (see Fig.~\ref{fig2}(d)). For
$\lambda$ = 0.7, 1 and 1.5, the optimal value of $\langle k \rangle$
is 20, 12 and 8 respectively (see Fig.~\ref{fig2}(e)). However, the
optimal value of $\langle k \rangle$ appears to be a non-monotonic
function of the rationality parameter $\beta$. For $\beta$ = 0.2, 1
and 10, the optimal value of $\langle k \rangle$ is 9, 8 and 9
respectively, as shown in Fig.~\ref{fig2}(f).

\begin{figure}
\begin{center}
 \scalebox{0.4}[0.4]{\includegraphics{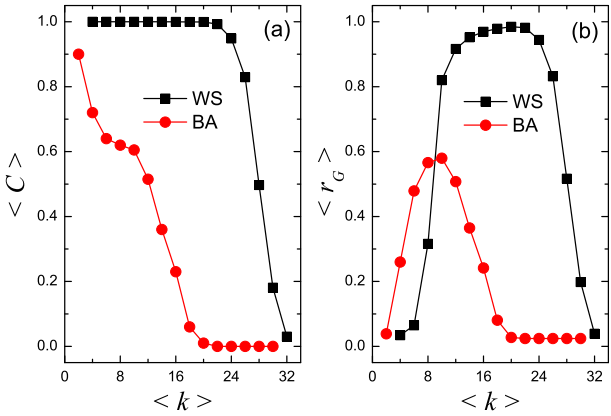}} \caption{(Color online)
(a) The average level of cooperation $\langle C \rangle$ and (b) the
average level of synchronization $\langle r_{G} \rangle$ as a
function of the average degree $\langle k \rangle$ for WS and BA
networks. The coupling strength $\lambda=1$, the relative cost
$\alpha=0.07$ and the rationality parameter $\beta=1$. In WS
networks, the rewiring probability is set to be 0.2.} \label{fig3.0}
\end{center}
\end{figure}

In the case of very small $\langle k\rangle$, the influence of
neighbors is so weak that oscillators are inclined to rotate at
their own frequencies. According to Eq.~(\ref{8}), we easily obtain
$\langle k \rangle/\langle k^{2} \rangle=1/(\langle k\rangle+1)$ for
ER random graphs. Thus the critical coupling strength arousing the
onset of a coherent state increases as the average degree $\langle k
\rangle$ decreases, indicating that one must strengthen the coupling
in order to reach the global synchronization for small $\langle k
\rangle$. In fact, small $\langle k\rangle$ not only hinders
synchronization in Kuramoto model but also impedes consensus in
opinion dynamics~\cite{opinion}. On the other hand, in the case of
very large $\langle k \rangle$, the system becomes close to the
well-mixed scenario. Due to the costly interaction, the payoffs of
cooperators are lower than those of defectors when $\langle k
\rangle$ is very large. In this case, the extinction of cooperators
is inevitable. The The full defection will lead to an incoherent
state of the system since all agents do not interact with neighbors
and rotate with their natural frequencies. The extinction of
cooperators for the case of large $\langle k \rangle$ has also been
found in the prisoner's dilemma game~\cite{pdg}. Combining the
discussion of the two limits of $\langle k \rangle$, highest
synchronization level should be realized for some intermediate
values of $\langle k\rangle$.

In the above studies, we use ER random graphs. In fact, qualitative
results remain unchanged for other kinds of networks including
Watts-Strogatz (WS) small-world networks~\cite{ws} and
Barab\'{a}si-Albert (BA) scale-free networks~\cite{ba}. From
Fig.~\ref{fig3.0}, one can see that for WS or BA networks, the
cooperation level $\langle C \rangle$ decreases as the average
degree $\langle k \rangle$ increases while the synchronization level
$\langle r_{G} \rangle$ is maximized at a moderate value of $\langle
k \rangle$.

\subsection{Effects of the rationality parameter $\beta$}

\begin{figure}
\begin{center}
 \scalebox{0.42}[0.42]{\includegraphics{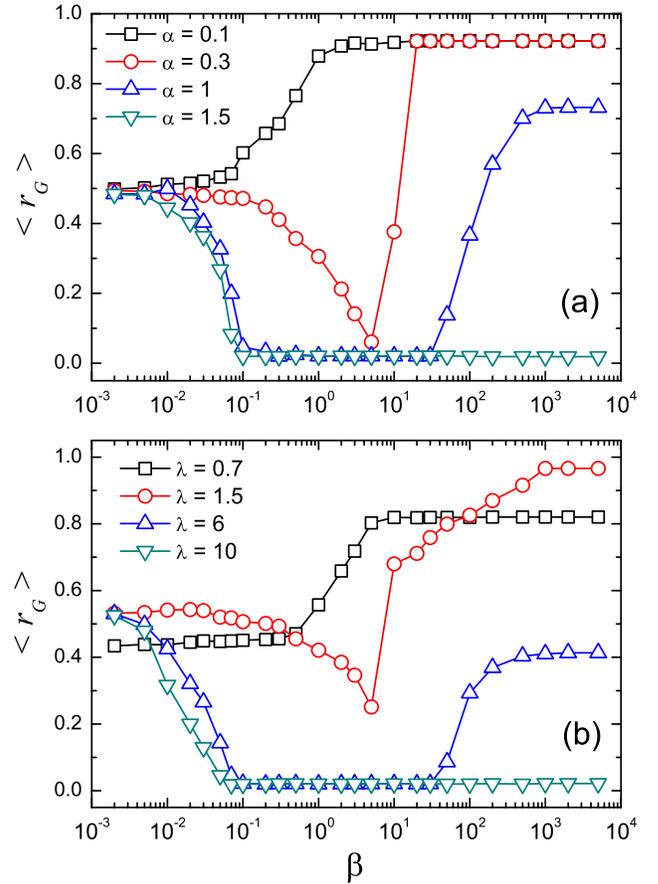}} \caption{(Color online) The average level of synchronization $\langle r_{G} \rangle$ as
 a function of the rationality parameter $\beta$ for different values of (a) the
relative cost $\alpha$ and (b) the coupling strength $\lambda$,
respectively. The average degree of the network is $\langle k
\rangle=8$. For (a), the coupling strength $\lambda=1$. For (b), the
relative cost $\alpha=0.2$.} \label{fig3}
\end{center}
\end{figure}

In the previous work~\cite{prl}, the rationality parameter $\beta$
is set to be 1. However, studies on other evolutionary games have
shown that, the cooperation level is largely affected by the
rationality parameter~\cite{noise1,noise2,noise3,noise4}.
Figure~\ref{fig3} shows the average level of synchronization
$\langle r_{G} \rangle$ as a function of the rationality parameter
$\beta$ for different values of the relative cost $\alpha$ and the
coupling strength $\lambda$. One can see that for small values of
$\alpha$ or $\lambda$ (e.g., $\alpha=0.1$ or  $\lambda=0.7$),
$\langle r_{G} \rangle$ increases with $\beta$. On the other hand,
for large values of $\alpha$ or $\lambda$ (e.g., $\alpha=1.5$ or
$\lambda=10$), $\langle r_{G} \rangle$ decreases as $\beta$
increases. For moderate values of $\alpha$ and $\lambda$, we notice
that $\langle r_{G}\rangle$ is minimized at the middle values of
$\beta$. The dependence of the average cooperation level $\langle C
\rangle$ on $\beta$ also obeys the above rule (results are not shown
here).

\begin{figure}
\begin{center}
 \scalebox{0.4}[0.4]{\includegraphics{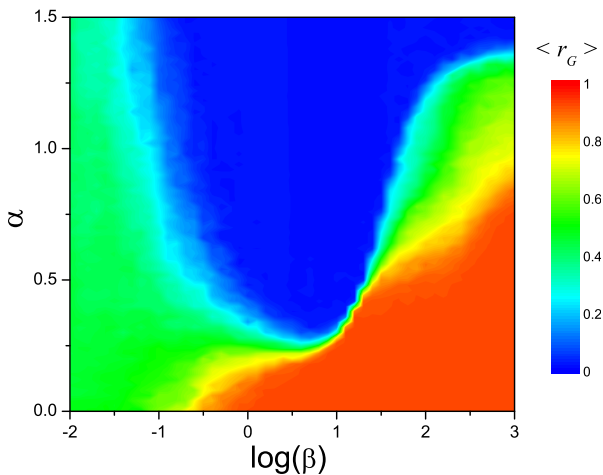}} \caption{(Color online) Color coded map of the average level of synchronization $\langle r_{G} \rangle$
in the parameter plane ($\beta$, $\alpha$). The average degree of
the network $\langle k \rangle=8$ and the coupling strength
$\lambda=1$.} \label{fig4}
\end{center}
\end{figure}

In Fig.~\ref{fig4}, we plot a color coded map of $\langle r_{G}
\rangle$ in the parameter plane ($\beta$, $\alpha$) by setting
$\langle k \rangle=8$ and $\lambda=1$. We find that, for
$\alpha<0.25$, $\langle r_{G} \rangle$ increases with $\beta$. For
$0.25<\alpha<1.4$, $\langle r_{G} \rangle$ is minimized at the
middle values of $\beta$. For $\alpha>1.4$, $\langle r_{G} \rangle$
decreases as $\beta$ increases.

It has been known that the formation of clusters plays an important
role in the evolutionary games~\cite{cluster1,cluster2}. A
cooperator (defector) cluster is a connected component fully
composed of cooperators (defectors). For small values of $\alpha$ or
$\lambda$, the interaction cost is much smaller than the benefits
due to mutual synchronization, leading to high payoffs of agents
inside cooperator clusters. The payoffs of agents inside defector
clusters are usually low since they change phases independently (no
local synchronization is expected). Thus for small values of
$\alpha$ or $\lambda$, the expansion of cooperator clusters becomes
easier as agents are more rational (i.e., the increase of $\beta$).
On the other hand, for very large values of $\alpha$ or $\lambda$,
the interaction cost is so large that cooperators cannot form stable
clusters to resist the invasion of defectors. In this case, the
decrease of $\beta$ can reduce the probability that a cooperator is
replaced by a defector. Note that $\beta=0$ corresponds to a random
strategy update, which is conceptually similar to a neutral drift of
the voter model~\cite{voter}. For $\beta=0$, the system has the
equal probability to terminate in the full cooperation or the full
defection, leading to the average cooperation level $\langle C
\rangle=0.5$.

\begin{figure}
\begin{center}
 \scalebox{0.42}[0.42]{\includegraphics{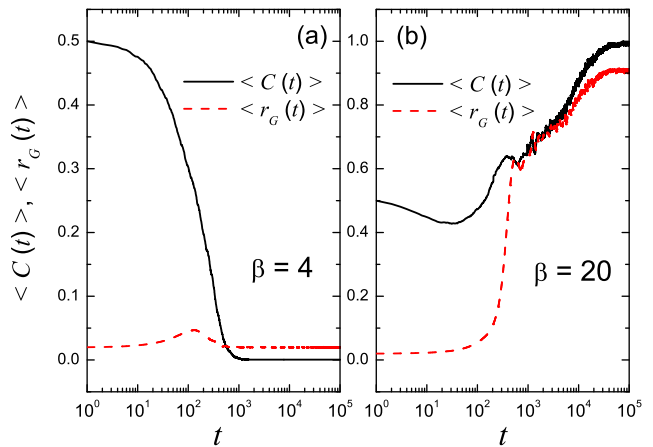}} \caption{(Color online) Time series of the
cooperation (synchronization) level $\langle C (t) \rangle$
($\langle r_{G} (t) \rangle$) for (a) $\beta=4$ and (b) $\beta=20$
respectively. The average degree of the network $\langle k
\rangle=8$, the relative cost $\alpha=0.4$ and the coupling strength
$\lambda=1$.} \label{fig5}
\end{center}
\end{figure}

To understand the nonmonotonic behavior appearing in the moderate
region of $\alpha$ and $\lambda$, we study the time evolution of the
cooperation level $\langle C (t) \rangle$ (synchronization level
$\langle r_{G} (t) \rangle$) for different values of the rationality
parameter $\beta$. From Fig.~\ref{fig5}(a), we see that, for
$\beta=4$, $\langle C (t) \rangle$ decreases to 0 as time evolves,
meanwhile, $\langle r_{G} (t) \rangle$ keeps close to 0 but slightly
peaks at about $t=100$. For $\beta=20$, $\langle C (t) \rangle$
initially decreases and then increases to 1, meanwhile, $\langle
r_{G} (t) \rangle$ gradually increases to about 0.9 (see
Fig.~\ref{fig5}(b)). From Fig.~\ref{fig5}, one can find that
cooperator clusters continuously shrink and finally disappear in the
case of $\beta=4$. However, for a large value of $\beta$ (e.g.,
$\beta=20$), cooperator clusters can survive after the initial
invasion of defectors. Note that the change of strategies happens on
the border that separates clusters of cooperators and defectors. For
a smaller value of $\beta$, cooperators along the border are more
likely to be replaced by neighboring defectors, leading to the
instability of cooperator clusters. However, for a very large value
of $\beta$, cooperators along the border cannot be invaded once
their payoffs are higher than those of neighboring defectors.
Cooperators inside a stable cluster can gain more and more benefit
from the mutual synchronization and the interaction cost continually
decreases as time evolves. From Fig.~\ref{fig5}(b), we observe that
$\langle r_{G} (t) \rangle$ rapidly increases from 0.05 to 0.6 while
$\langle C (t) \rangle$ keeps around 0.6 during $200<t<500$. In this
period, cooperators get more and more synchronized, leading to the
latter growth of cooperator clusters. Summarizing, for moderate
interaction cost, sufficient high extent of rationality helps
cooperators to survive and form tiny clusters to compete with
defectors in the initial stage, and then enables them to strike
back, leading to the high level of cooperation and synchronization
in the later stage. Moderate rationality is, however, lack of the
ability to stabilize the cluster of cooperators, resulting in the
lowest level of cooperation and synchronization. It was worth noting
that such nonmonotonic phenomenon, i.e., the cooperation level
reaches the minimum at a moderate value of the rationality
parameter, has also been observed in the spatial public goods
game~\cite{pgg} and snowdrift game~\cite{sg}.

\begin{figure}
\begin{center}
 \scalebox{0.4}[0.4]{\includegraphics{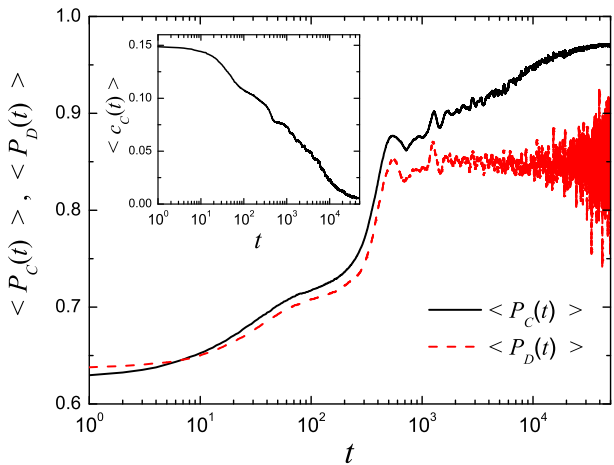}} \caption{(Color online) Time series of the
average payoff of cooperators $\langle P_{C} (t) \rangle$ and
defectors $\langle P_{D} (t) \rangle$. The inset shows the average
cost of cooperators as time evolves. The average degree of the
network $\langle k \rangle=8$, the relative cost $\alpha=0.4$, the
coupling strength $\lambda=1$ and the rationality parameter
$\beta=20$.} \label{fig6}
\end{center}
\end{figure}

Figure~\ref{fig6} shows time series of the average payoff of
cooperators $\langle P_{C} (t) \rangle$ and defectors $\langle P_{D}
(t) \rangle$ for the case of $\beta=20$. One can see that initially
$\langle P_{C} (t) \rangle$ is lower that $\langle P_{D} (t)
\rangle$. When $t>10$, $\langle P_{C} (t)\rangle$ gradually exceeds
$\langle P_{D} (t)\rangle$. From the inset of Fig.~\ref{fig5}, we
observe that the average cost of cooperators $\langle c_{C} (t)
\rangle$ decreases as time evolves. In the late stage of evolution,
the system is almost occupied by cooperators and only a few
defectors survive. The phases of these cooperators are almost the
same while defectors change their phases independently. Thus the
payoff of a defector fluctuates greatly in the late stage of
evolution. A defector gain a high (low) payoff when its phase is the
same (different) with those of cooperators.

\section{Conclusions}

In conclusion, we have studied how the average interaction degree
and the extent of rationality affect the coevolution of cooperation
and synchronization. Our main findings can be summarized as follows.
(i) The cooperation level is positively correlated with the
synchronization level, that is, more cooperators will promote
synchronization. (ii) Both of the cooperation and synchronization
levels decrease as the relative cost increases. (iii) Both of the
cooperation and synchronization levels are maximized in the middle
range of the coupling strength. (iv) The cooperation level decreases
as the average degree of the network increases. But the
synchronization level is maximized at a moderate value of the
average degree. (v) The extent of rationality plays a nontrivial
role in the coupling dynamics. For small (large) values of the
relative cost or the coupling strength, both the cooperation and
synchronization levels increase (decrease) as the rationality
parameter increases. For moderate values of the relative cost and
the coupling strength, however, both the cooperation and
synchronization levels are minimized at a middle value of the
rationality parameter.

Our results offer a deeper understanding of the interplay of
synchronization dynamics and game theory. There remains a number of
open questions in EKG. For example, how the clustering coefficient
and degree correlations of the underlying interaction network affect
the synchronization? What would happen if we use other rules of
strategy updating in EKG? We hope our work can stimulate more
researchers into the study of coevolution of synchronization and
cooperation.

\begin{acknowledgments}

This work was supported by the National Science Foundation of China
under Grant Nos.~61773121, 61403083, 11475074, and 11575072, and the
Qishan scholar research fund of Fuzhou University

\end{acknowledgments}

\end{document}